\documentclass[%
 aip,
 amsmath,amssymb,
 reprint,%
]{revtex4-1}

\usepackage{graphicx}
\usepackage{dcolumn}
\usepackage{bm}

\usepackage[utf8]{inputenc}
\usepackage[T1]{fontenc}
\usepackage{mathptmx}
\usepackage{etoolbox}
\usepackage{subcaption}
\usepackage{caption}

\makeatletter
\def\@email#1#2{%
 \endgroup
 \patchcmd{\titleblock@produce}
  {\frontmatter@RRAPformat}
  {\frontmatter@RRAPformat{\produce@RRAP{*#1\href{mailto:#2}{#2}}}\frontmatter@RRAPformat}
  {}{}
}%
\makeatother
\begin{document}

\preprint{AIP/123-QED}

\title[Coherent Organizational States in Turbulent Pipe Flow at moderate Reynolds numbers]{Coherent Organizational States in Turbulent Pipe Flow at moderate Reynolds numbers}
\author{R. Jäckel}
  \email{r.jackel@mecanica.coppe.ufrj.br}

\affiliation{ 
Mechanical Engineering Program (PEM/COPPE/UFRJ), Federal University of Rio de Janeiro, Rio de Janeiro, Brazil
}%
\affiliation{ 
Interdisciplinary Center for Fluid Dynamics, Federal University of Rio de Janeiro, Rio de Janeiro, Brazil
}%

\author{B. Magacho}%

\affiliation{%
Instituto de Física, Federal University of Rio de Janeiro, Rio de Janeiro, Brazil
}%
\affiliation{ 
Interdisciplinary Center for Fluid Dynamics, Federal University of Rio de Janeiro, Rio de Janeiro, Brazil
}%

\author{B. E. Owolabi}

\affiliation{ 
Mechanical Engineering Program (PEM/COPPE/UFRJ), Federal University of Rio de Janeiro, Rio de Janeiro, Brazil
}%
\affiliation{ 
Interdisciplinary Center for Fluid Dynamics, Federal University of Rio de Janeiro, Rio de Janeiro, Brazil
}%

\author{L. Moriconi}%

\affiliation{%
Instituto de Física, Federal University of Rio de Janeiro, Rio de Janeiro, Brazil
}%
\affiliation{ 
Interdisciplinary Center for Fluid Dynamics, Federal University of Rio de Janeiro, Rio de Janeiro, Brazil
}%

\author{D. J. C. Dennis}
\affiliation{ 
Interdisciplinary Center for Fluid Dynamics, Federal University of Rio de Janeiro, Rio de Janeiro, Brazil
}%

\author{J. B. R. Loureiro}
\affiliation{ 
Mechanical Engineering Program (PEM/COPPE/UFRJ), Federal University of Rio de Janeiro, Rio de Janeiro, Brazil
}%
\affiliation{ 
Interdisciplinary Center for Fluid Dynamics, Federal University of Rio de Janeiro, Rio de Janeiro, Brazil
}%

\date{\today}

\begin{abstract}

Turbulent pipe flow is still an essentially open area of research, boosted in the last two decades by considerable progress achieved both on the experimental and numerical frontiers, mainly related to the identification and characterization of coherent structures as basic building blocks of turbulence. It has been a challenging task, however, to detect and visualize these coherent states. 
We address, by means of stereoscopic particle image velocimetry, that issue with the help of a large diameter (6 inches) pipe loop, which allowed us to probe for coherent states at various moderate Reynolds numbers (5300 < Re < 29000)). Although these states have been observed at flow regimes around laminar-turbulent transition (Re $\approx$  2300) and also at high Reynolds number pipe flow (Re $\approx$ 35000), at moderate Reynolds numbers their existence had not been observed yet by experiment. By conditionally averaging the flow fields with respect to their dominant azimuthal wavenumber of streamwise velocity streaks, we have been able to uncover the existence of ten well-defined coherent flow patterns.  
It turns out, as a remarkable phenomenon, that their occurrence probabilities and the total number of dominant modes do not essentially change as the Reynolds number is varied. Their occurrence probabilities are noted to be reasonably well described by a Poisson distribution, which suggests that low-speed streaks are created as a Poisson process on the pipe circular geometry.
\end{abstract}


\maketitle

\section{\label{sec:level1}Introduction}

Turbulent structures in pipe flows have been a subject of great interest in fluid dynamics since the very first pioneering experiments of Reynolds in 1883 \cite{Reynolds1883}. Until recently, research in turbulence was mainly focused on the statistical perspective of distinct flow features such as the statistical distribution of flow variables\cite{Eggels1994, DenToonder1997, Wallace2016}, turbulent energy spectra \cite{Tsuji_2004, Meyers_2008}, or RANS modeling \cite{Jin_Long_Wu_2018}. A relatively new approach, on the other hand, often-referred to as dynamical systems viewpoint emerged in the last two decades due to considerable progress achieved both on the experimental and numerical frontiers. This essentially open area of research is related to the identification and characterization of coherent structures \cite{Hussain_1983,Grinstein_1996,Chrisohoides_2003,Khan_2020}. 

The term coherent (from lat. Cohaerens – consistency) emphasizes on the understanding that turbulence, contrary to earlier assumptions, is no longer an example of chaos, but rather a superposition of canonical building blocks of motion with inherent patterns of spatial and temporal consistency. A better understanding of these building blocks has been the motivation for hot contemporary debates and the elaboration of improved experimental set-ups \cite{Hof2011}. The near-wall production and complex dynamics of evolving coherent structures (e.g., turbulent puffs, quasi-streamwise and hairpin vortices, etc.) have been the fundamental keywords in these developments \cite{Dennis2015}.

It is clear, however, that a gap in the literature persists, related to the visualization of near-wall coherent structures in pipe flows, in order to see how they can validate, refine or even suggest alternative perspectives to the ongoing scientific discussions. It is well known that turbulent statistics for the logarithmic region below Reynolds numbers of 25000 lack universality due to their Reynolds number dependence, a phenomenon often referred to as Reynolds number effect \cite{DenToonder1997,Zhou2021,Fischer_2001}. It is still under debate if this unique behavior is also shown by coherent structures. 
In this work, we provide a new step towards closing these gaps, exploring, with the help of Stereoscopic Particle Image Velocimetry (SPIV), the intriguing patterns of near-wall coherent structures associated with turbulent regimes in pipe flow, applying methodological lines similar to those applied by Hof et al.\cite{Hof2004}, Schneider et al. \cite{Schneider2007} and Dennis and Sogaro \cite{Dennis2008}. 

Hof et al. \cite{Hof2004} as one of the first presented a proper visualization of traveling waves as coherent structures in pipe flow at Reynolds numbers close to the laminar-turbulent transition by means of SPIV experiments. The observed structures showed azimuthal patterns of high-speed streaks close to the wall and low-speed streaks closer to the pipe centre. 

Not very long after, Schneider et al. \cite{Schneider2007} , by means of numerical simulation performed further investigations. They established a new approach for the structure identification, which allowed them to uncover a huge number of different coherent states together with their statistical features. These authors furthermore suggested that the transition dynamics could be modeled as a Markovian stochastic process, a phenomenological point that has been addressed in the recent literature \cite{Jäckel_2023}.   

Both Hof et al. \cite{Hof2004}  and Schneider et al.\cite{Schneider2007}  interpreted traveling waves as phenomena related to laminar-turbulent transition. This assumption was called into question by Dennis and Sogaro’s \cite{Dennis2008} SPIV experiments in pipe flow at a highly turbulent regime of Re = 35000. They showed that their flow was also organized into different coherent states, which bear a striking resemblance to travelling wave solutions observed until then only at lower Reynolds numbers, with the propensity for switching from one mode to another. 

In this study, we further investigate the turbulent states by closing the huge gap between these coherent states observed in laminar-turbulent transition by Hof et al. \cite{Hof2004}  and Schneider et al.\cite{Schneider2007} , and those observed at relatively high Reynolds number by Dennis and Sogaro \cite{Dennis2008}  in order to get a clearer picture of how these states evolve with the Reynolds number. 

The experimental setup, to be described in the next section, allows us to get a deep insight into the boundary layer in conditions of moderate Reynolds number turbulence. Also, we explain in detail the methods applied to detect and visualize the coherent states.

In Sec. 3, we present the results of our work in a twofold manner; qualitatively, by visualization of the conditionally averaged cross-stream patterns associated with the dominant coherent states and their organization along the mean flow direction, and quantitatively by showing interesting statistical features of the occurrence probabilities of these states.   

Finally, in Sec. 4, we summarize and discuss the main ideas of our work and give an outlook on future work required to further improve our understanding regarding the nature of coherent states.

The experimental results and statistical analysis will uncover turbulent dynamics so far unexplored by providing unique insight into turbulent pipe flow regarding its phenomenology \textit{vis-à-vis} with statistical features which are highly relevant to better understand, predict and model turbulence. 


\section{Materials and Methods}

\subsection{\label{sec:level2}Experimental setup}
The experiment was performed in a flow loop specially designed for the research on wall turbulence and coherent structures. The flow loop consists of a horizontal 6-inch diameter, 10 meters long pipe, operating in a closed system. By means of a progressive cavity pump, water is driven from a large reservoir through a Coriolis flow meter before entering the pipe. All components are connected by a flexible 2-inch rubber hose, which further serves as a pulsation damper. A settling chamber consisting of a diffuser cone with a 6-degree angle and a 1:3 aspect ratio followed by a honeycomb and a set of screens was installed to reduce eddies and swirling motions before the flow enters the pipe. We estimate the hydrodynamic entrance length for turbulent pipe flow by means of Eq. \ref{eq:entrance_length} (Ref. \cite{Bhatti1987}):

\begin{eqnarray}
L_{H} = 1.359 D (Re)^{1/4}    
\label{eq:entrance_length}
\end{eqnarray}

\noindent
and expect the flow to be fully developed long before entering the observation section located at 5.8 meters downstream for the Reynolds number with the longest entrance length studied in this work, namely Re = 29000 ($L_{H} = 2.73$ m).

\begin{figure*}
\centering
\includegraphics[width=0.9\textwidth]{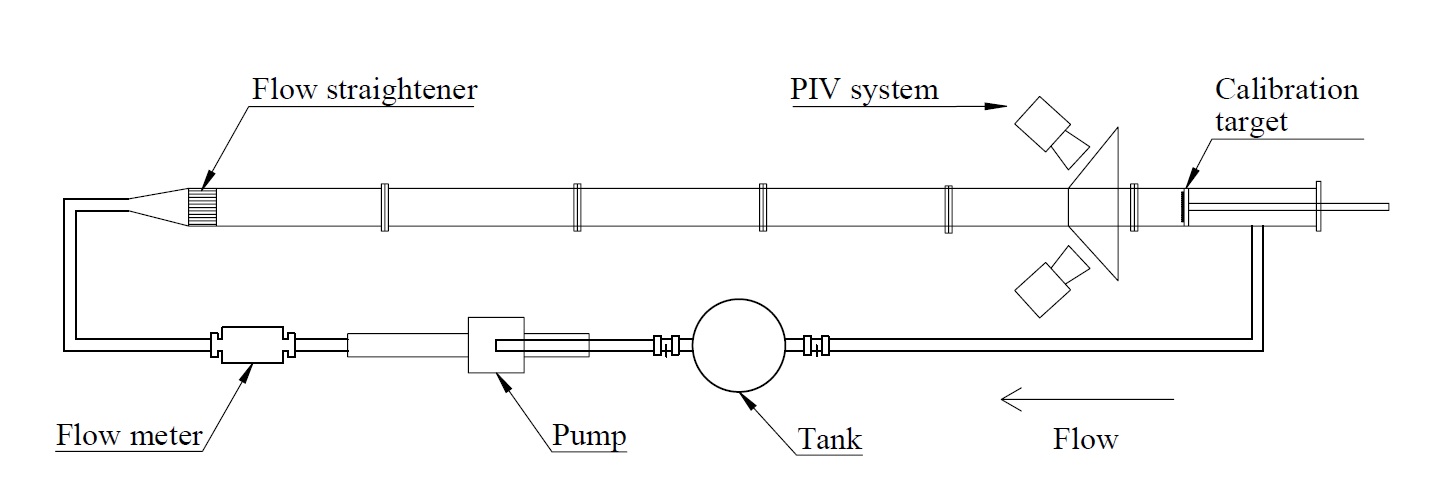}
\caption{\label{fig:rig}Experimental setup (not to scale) of the pipe rig with SPIV system. The flow direction is clockwise.}

\end{figure*}

Our experiment is based on a time-resolved Stereo PIV (SPIV) setup with two high-speed CMOS cameras (Phantom Speed-sense M310), arranged horizontally at an angle of 45 degrees to the pipe centerline aiming downstream in order to capture a transversal plane of the flow as shown in Fig. \ref{fig:rig}.

The water-filled trapezoidal section was used to minimize optical distortions caused by the pipe curvature.
A two-level 15.4 cm diameter calibration target, visible for each camera through the same angle of 45 degrees relative to the pipe, was moved into the measurement plane to calibrate the SPIV system by means of a long, pipe-centered traverse mechanism. After calibration, the target was moved downstream into a parking position located behind the pipe outlet in order to avoid any flow disturbance. The flow was seeded with silver-coated hollow glass spheres, neutrally buoyant with a mean size of 17 microns, which accurately follow, as tracers, turbulent fluctuations of the flow field. 
All Reynolds number measurements were acquired with a sampling frequency of 15 Hz and a considerable number of acquisitions, approximately 20000 captured vector fields for each run, which result in approximately 308, 628, 829, 1259, and 1492 pipe radii passing through the measurement plane for Re = 5300, 12000, 17800, 24400 and 29000, respectively. Each Reynolds number measurement was acquired in subsets of 2000 snapshots which were separated by time intervals of several minutes. We, therefore, consider these subsets statistically independent and the overall statistics of each Reynolds number set hardly to be distorted by any kind of Very Large Scale Motions (VLSMs) \cite{Hellström_2017}.

\begin{figure}
\centering
\includegraphics[width=0.46\textwidth]{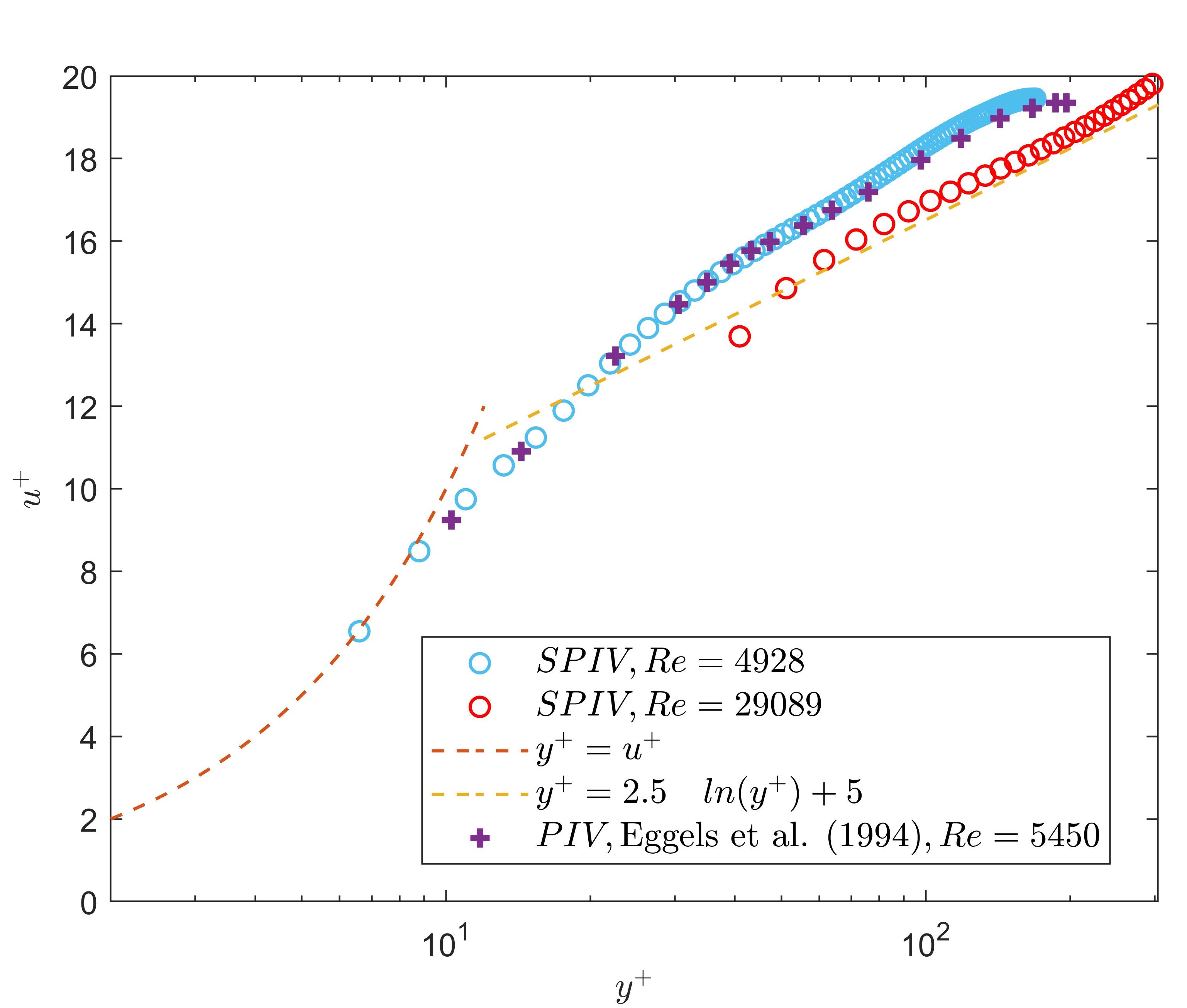}
\caption{\label{fig:log_profiles} Streamwise velocity profile in inner units at a Reynolds number of 4928 and 29089 obtained with SPIV, compared with the results of Eggels et al. \cite{Eggels1994}.}
\end{figure}

Within the lower and upper bounds of the Reynolds numbers
studied in this work, the turbulent statistics measured
with our SPIV setup show very good agreement with data
from DenToonder and Nieuwstadt \cite{DenToonder1997}  and Eggels et al. \cite{Eggels1994} and also
clearly shows the aforementioned Reynolds number effect
in the log region, as demonstrated in Fig. \ref{fig:log_profiles}.

\begin{figure}
\centering
\includegraphics[width=0.4\textwidth]{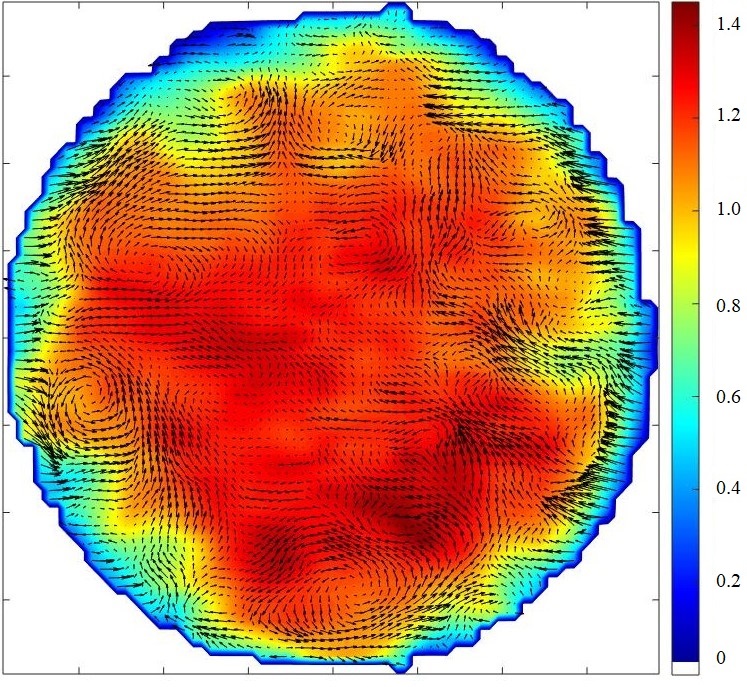}
\caption{\label{fig:cross_stream} Instantaneous vector field for the flow at Re = 24414. The color bar indicates the magnitude of the streamwise velocity component normalized by the bulk velocity.}
\end{figure}

Also, as demonstrated in Fig. \ref{fig:cross_stream}, the cross-stream vector fields we obtain with our measurement system are well-suited for the detection of near-wall structures, both for stream-wise streaks and also for in-plane motions like quasi-streamwise vortices (at least four of them can be detected directly by eye in this exemplary snapshot of Re = 24414). 

\subsection{\label{Detection}Detection and visualization of coherent states}

\begin{figure}
\centering
\includegraphics[width=0.5\textwidth]{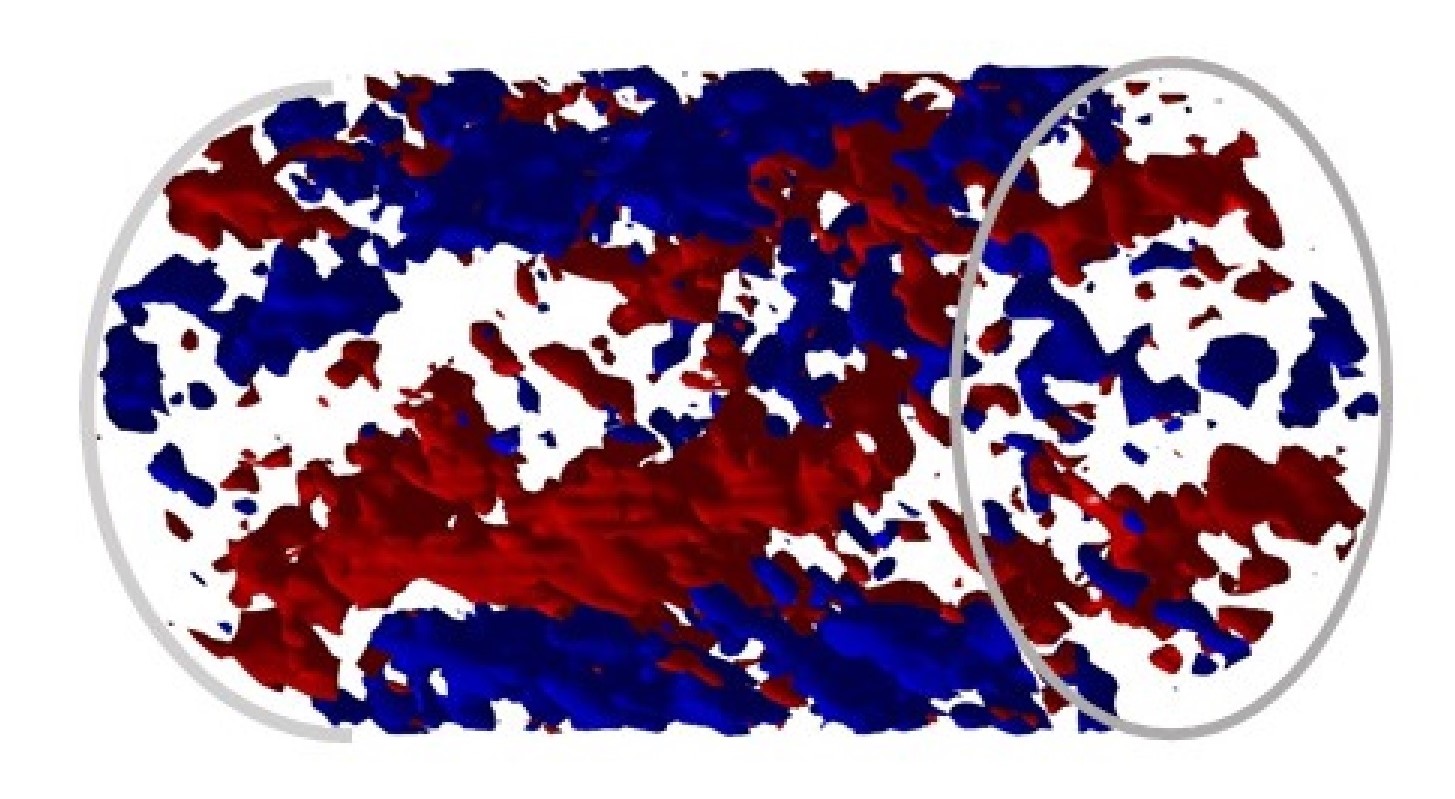}
\caption{\label{fig:stuctures_along_pipe} Velocity fluctuations along the mean flow direction (flow from left to right) at a Reynolds number of 5300. The red and blue iso-contours correspond to velocities 1.5 per cent above and below the mean velocity profile, respectively.}
\end{figure}

We base our coherent state detection on the appearance of positive and negative velocity fluctuations of the streamwise velocity components in each snapshot. These elongated, meandering regions of opposed fluctuations are advected along the mean flow direction, as seen in Fig. \ref{fig:stuctures_along_pipe} observed at a Reynolds number of Re = 5300.  

\begin{figure}
\centering
\includegraphics[width=0.49\textwidth]{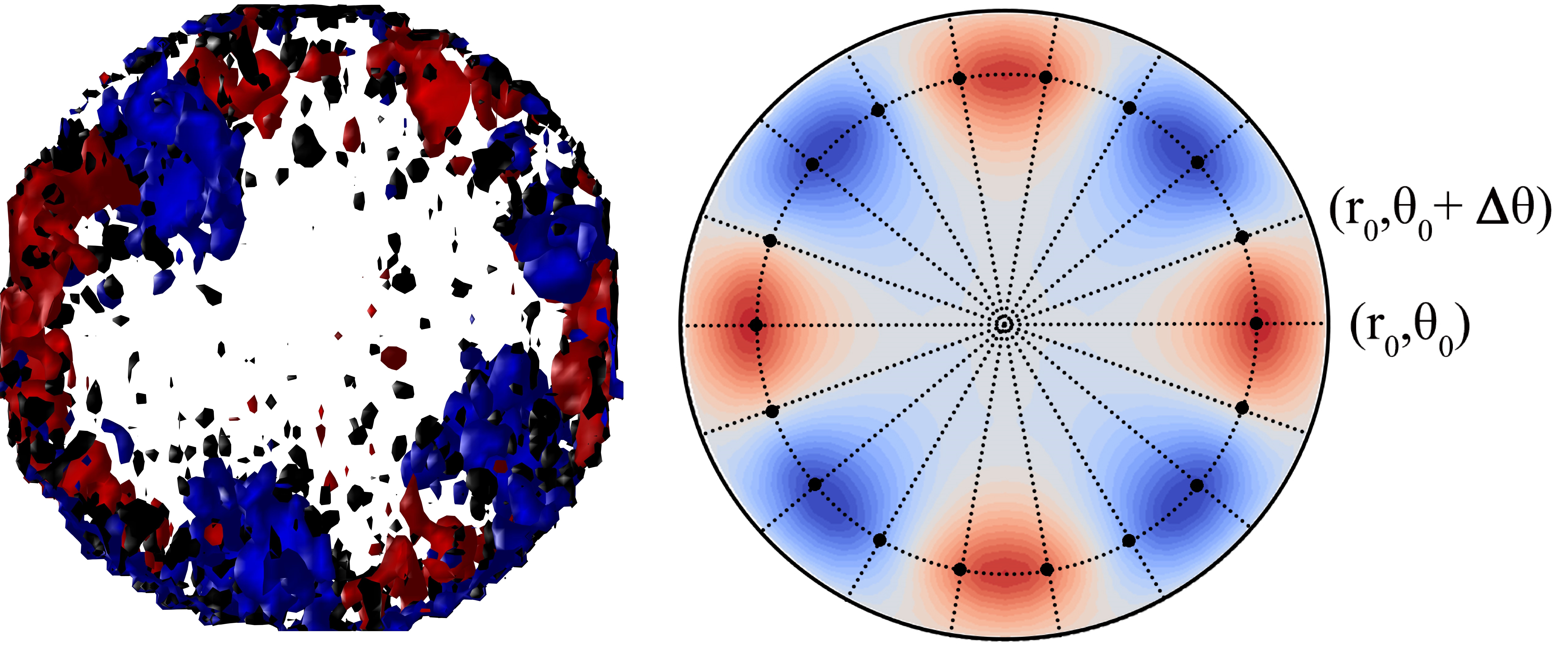}
\caption{\label{fig:stuctures_along_slice} Left: Instantaneous snapshot with an alternating pattern of streamwise velocity fluctuations.The red and blue iso-contours correspond to velocities 1.5 per cent above and below the mean velocity profile, respectively. Right: reference point ($r_{0}$, $\theta_{0}$) and azimuthal grid projected on an idealized pattern of spatial correlations for a wave number of 4. }
\end{figure}

\noindent
In a cross-stream slice, these fluctuations appear as an alternating pattern of positive and negative fluctuations as seen in Fig.\ref{fig:stuctures_along_slice} (left), similar to the ones also observed by Dennis and Sogaro \cite{Dennis2008}.

\begin{figure}
\centering
\includegraphics[width=0.5\textwidth]{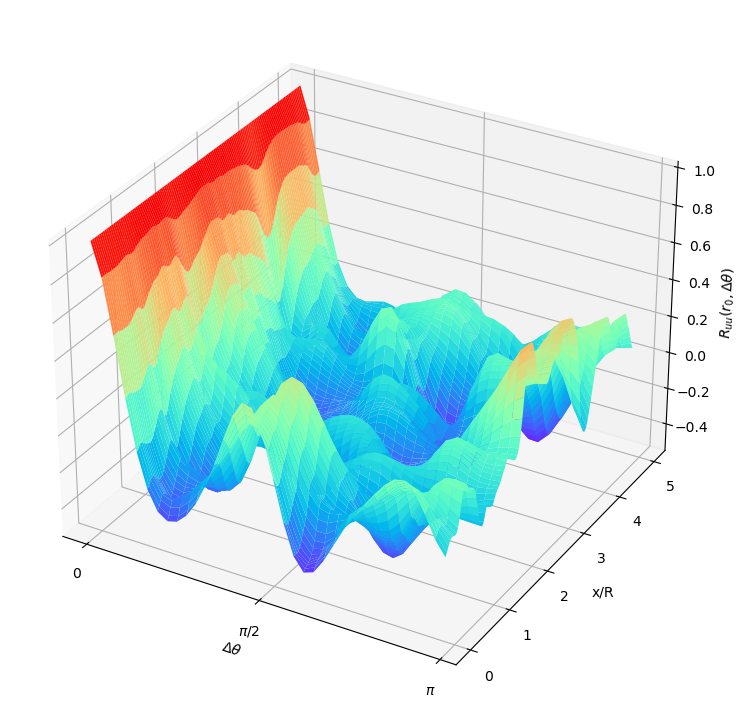}
\caption{\label{fig:azimuthal_correlation} Streamwise extent of the azimuthal correlation over a length of 5 radii, here for the Reynolds number of 5300. Positive peaks indicate a correlation, negative peaks anti-correlation. Note that for reasons of symmetry we only plot along an azimuth from zero to $\pi$. }
\end{figure}

Nonetheless, due to the high number of vector fields obtained for each Reynolds number set, the state detection together with the subsequent wave number assignment of the corresponding snapshots require an automated procedure. In the following, we present a procedure to reduce the complex pattern of a 3-dimensional flow field to only one parameter, i.e. the azimuthal wave number, to which each snapshot will be assigned subsequently.  
 	First, we define a reference point ($r_{0}$, $\theta_{0}$) and introduce the spatial correlation function between the reference point and equally distributed points located along the $r = r_{0}$ circumference with an azimuthal spacing of $\Delta\theta$ (see Fig. \ref{fig:stuctures_along_slice} (right)) by means of Eq. \ref{eq:spatial_correlation}:

\begin{eqnarray}
R_{uu}(r_{0}+\Delta r,\Delta \theta)= 	\frac{\langle u(r_{0}+\Delta r,\theta_{0}+\Delta\theta)~u(r_{0},\theta_{0})\rangle}{u^{2}_{rms}}       
\label{eq:spatial_correlation}
\end{eqnarray}

\noindent
with an azimuthal spacing of $\Delta\theta$ resulting in 72 equally spaced azimuthal points. The square brackets indicate an azimuthal average over the initial angles $ \theta_{0}$. 
If we limit Eq. \ref{eq:spatial_correlation}  to a fixed radius of interest, in this work $r_{0}=0.78R$, it turns into an azimuthal correlation function which will be used for the state detection. Fig. \ref{fig:azimuthal_correlation} shows an arbitrary section of the azimuthal correlation over a length of 5 radii, here for the Reynolds number of 5300, from zero to $\pi$.

Note that in the reference point itself, $\Delta_{\theta} =0$, the correlation is one, as by definition. 
Further, it is possible to observe how the azimuthal correlation function changes its number of peaks several times along this streamwise dimension.  
We obtain the corresponding azimuthal wave number by taking the highest value of the power of the Fast Fourier Transform (FFT) on the azimuthal correlation, which relates the flow field to a well-defined wavenumber-labeled state. In this way, all snapshots can be classified, according to their corresponding wavenumber subset.
 For the visualization of the spatial correlation not only above a single circumference line but on the entire cross-stream plane  ($C=C(r, \theta)$), we expand the azimuthal correlation along a radial grid with a spacing of 1 mm and plot the corresponding iso-surfaces of positive and negative correlation.     
In order to take advantage of the entire data set of flow fields obtained by our measurements, we apply a conditional average procedure in a twofold manner: The subsets of the iso-surface cross-stream plane correlations are averaged with regard to their allocated wavenumber. By means of this averaging procedure, we expect the patterns to smooth out and obtain figures comparable to the contours in Fig. \ref{fig:stuctures_along_slice} (right), in this case representative for the subset of azimuthal wave number $k_{\theta}=4$.  

For wall-bounded turbulence \cite{Dennis2015} and coherent states \cite{Dennis2008} it is known that regions of negative streamwise fluctuations are often related to different in-plane motions than positive streamwise fluctuations, namely $Q_2$ and $Q_4$ quadrant motions (also known as ejections and sweeps \cite{Wallace2016}). 
Therefore, for the flow field averaging, we further bifurcate our sampling condition by dividing the wavenumber subsets into these snapshots with a positive and those with a negative streamwise velocity fluctuation in the initial reference point.
To visualize the spatial distribution of the coherent states along the stream direction Taylor’s hypothesis \cite{Taylor1938} was applied.   
In the appendix, we present a flow diagram (Fig. \ref{fig:principal_steps}) which illustrates the principal steps of the procedure for the detection and visualization of coherent states in a concise manner.

\section{Results}

\subsubsection{\label{sec:level3}Flow patterns of coherent states}

\begin{figure*}
     \centering
    \begin{subfigure}[t]{0.26\textwidth}
        \raisebox{-\height}{\includegraphics[width=\textwidth]{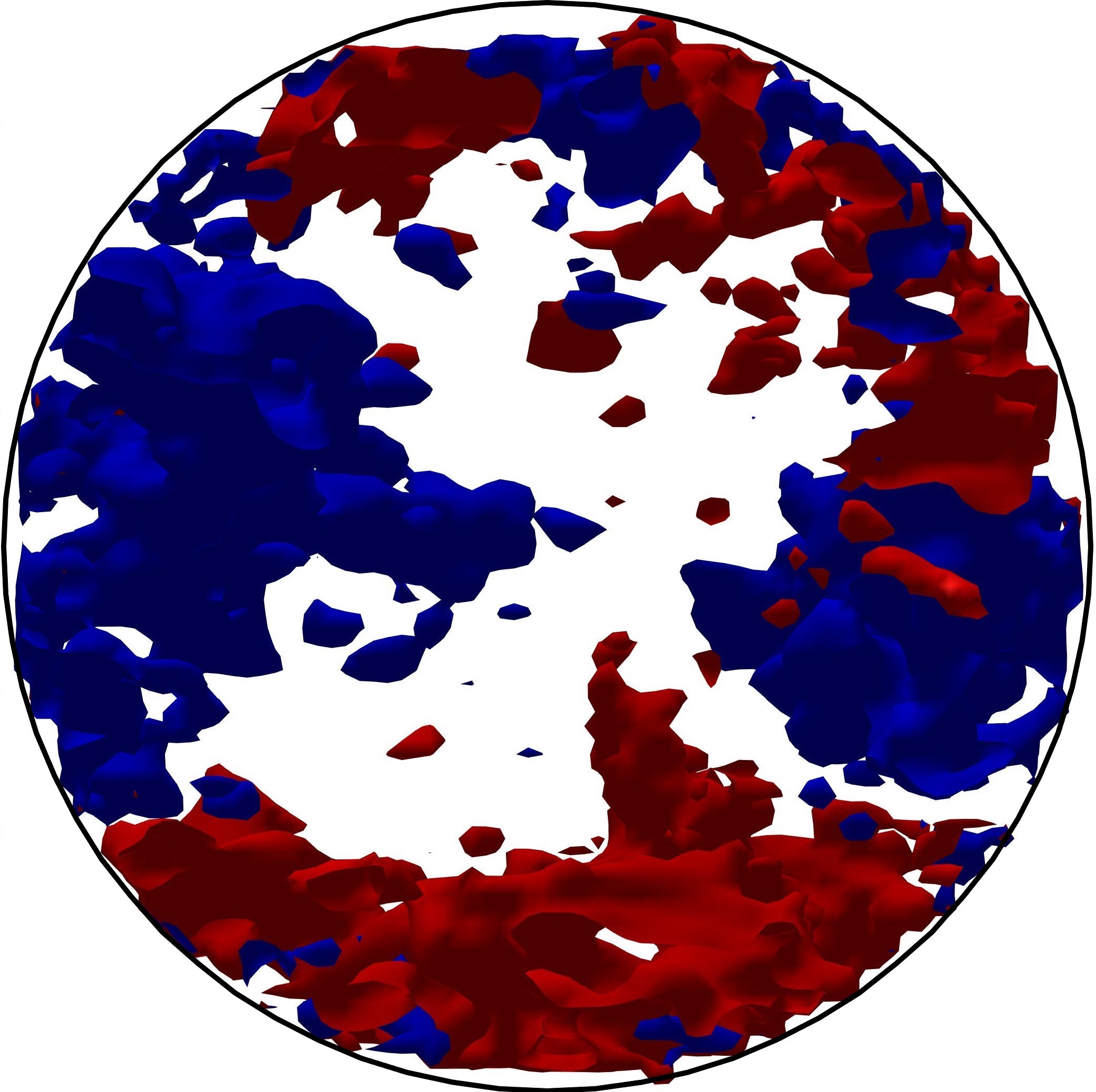}}
        \caption{$~k_{\theta}=2$}
    \end{subfigure}
    \hfill
    \begin{subfigure}[t]{0.26\textwidth}
        \raisebox{-\height}{\includegraphics[width=\textwidth]{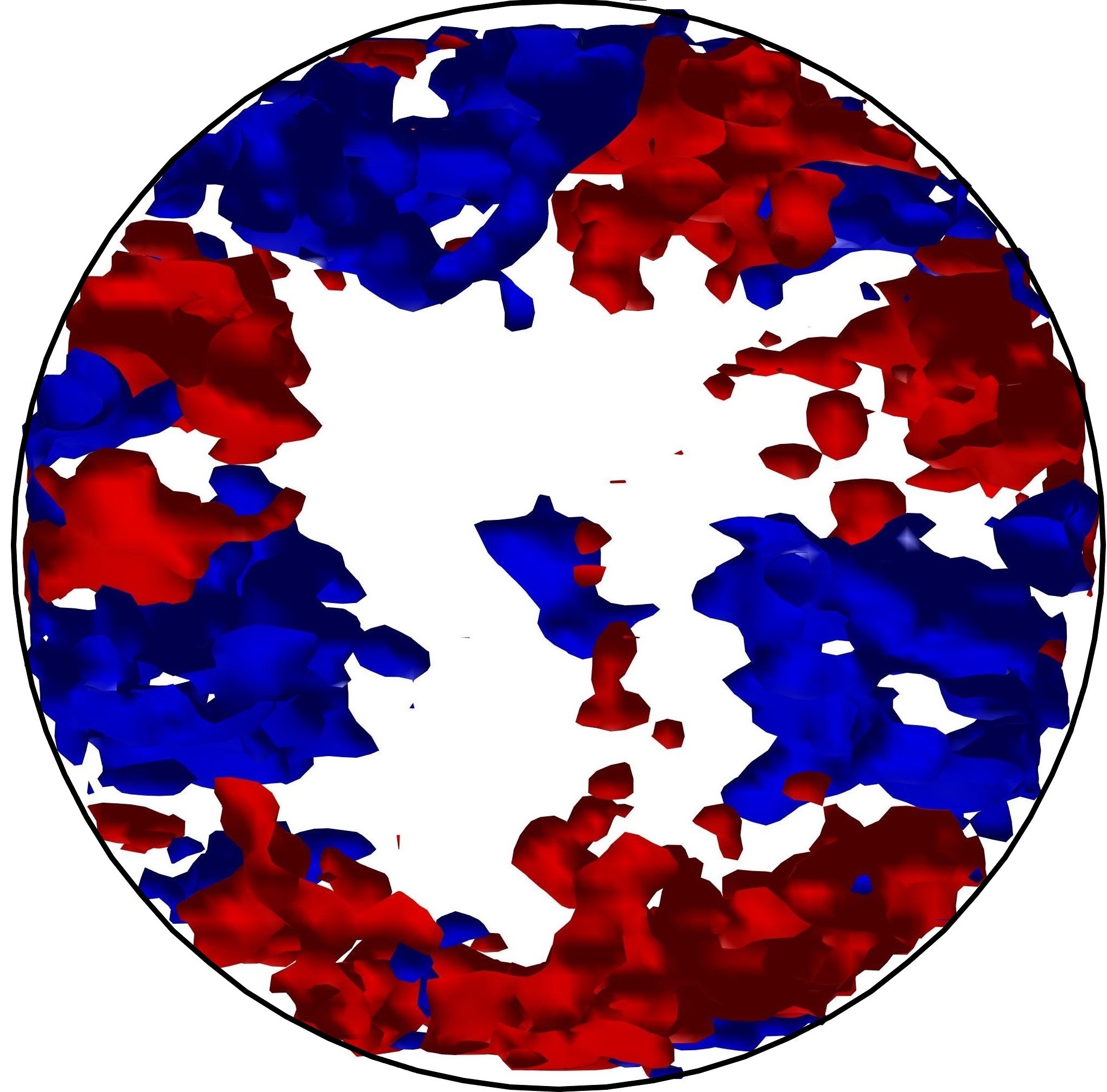}}
        \caption{$~k_{\theta}=3$}
    \end{subfigure}
        \hfill
\begin{subfigure}[t]{0.26\textwidth}
        \raisebox{-\height}{\includegraphics[width=\textwidth]{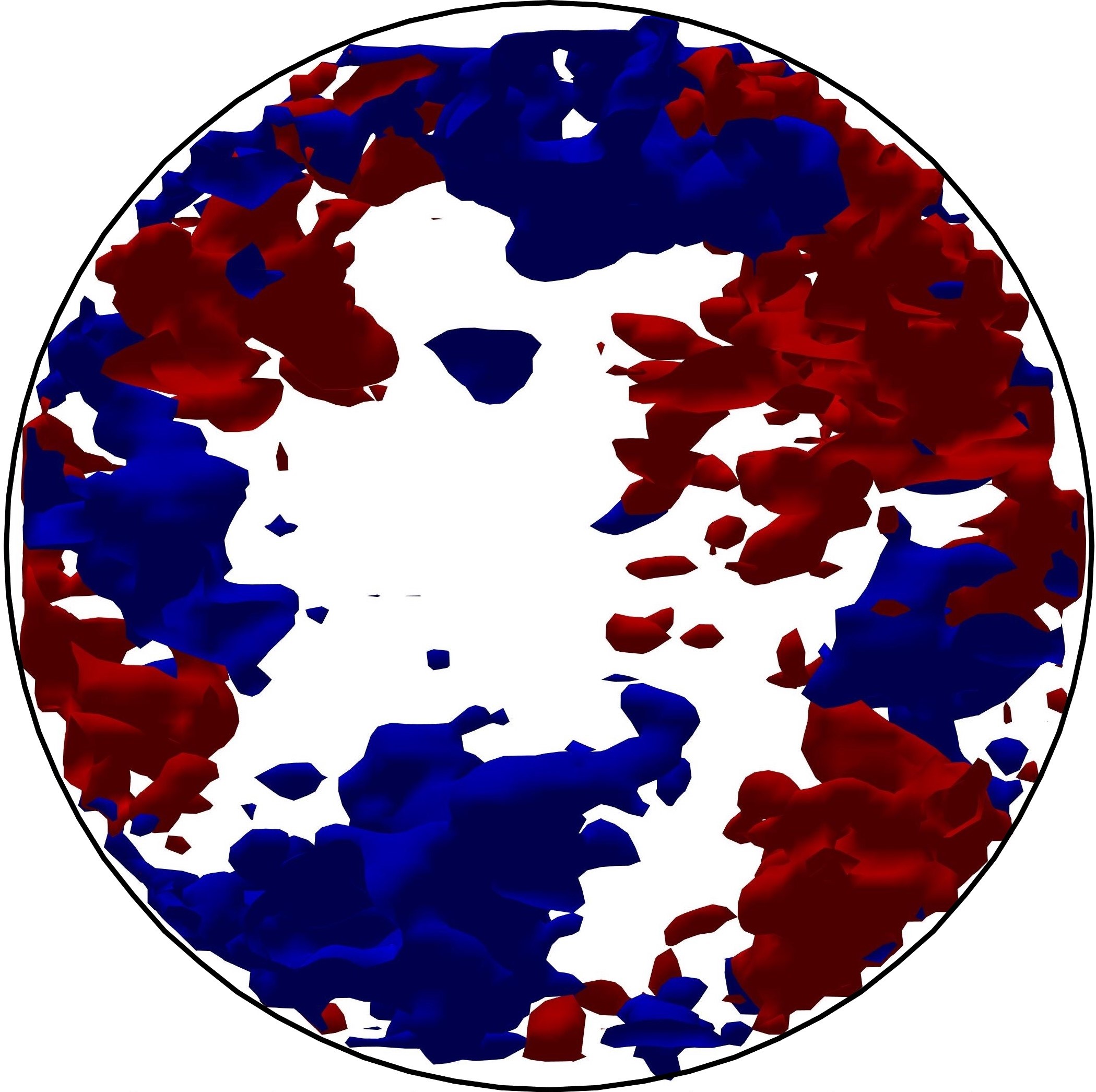}}
        \caption{$~k_{\theta}=4$}
    \end{subfigure}

\vspace{0.5cm}

 \begin{subfigure}[t]{0.26\textwidth}
        \raisebox{-\height}{\includegraphics[width=\textwidth]{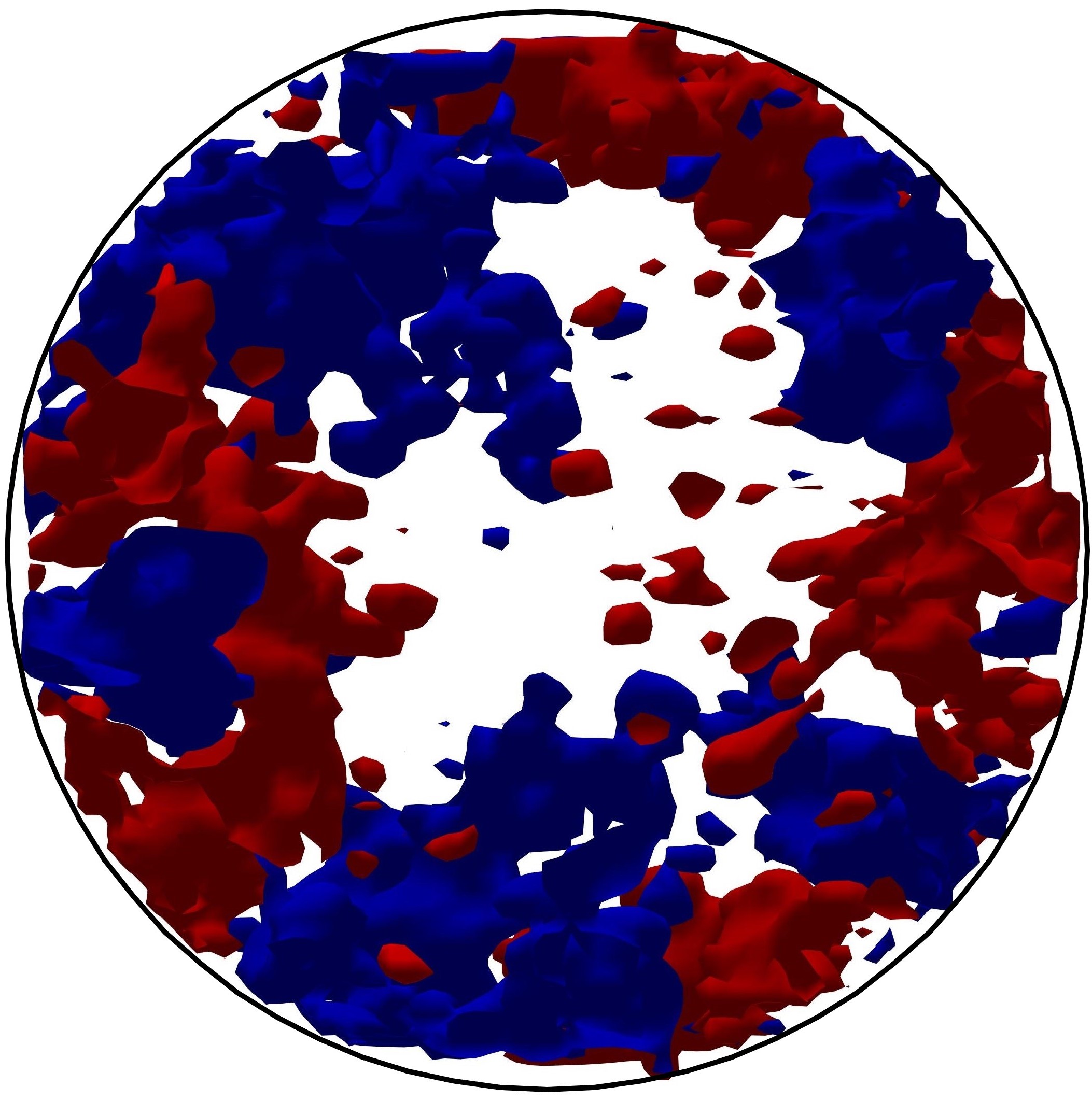}}
        \caption{$~k_{\theta}=5$}
    \end{subfigure}
    \hfill
    \begin{subfigure}[t]{0.26\textwidth}
        \raisebox{-\height}{\includegraphics[width=\textwidth]{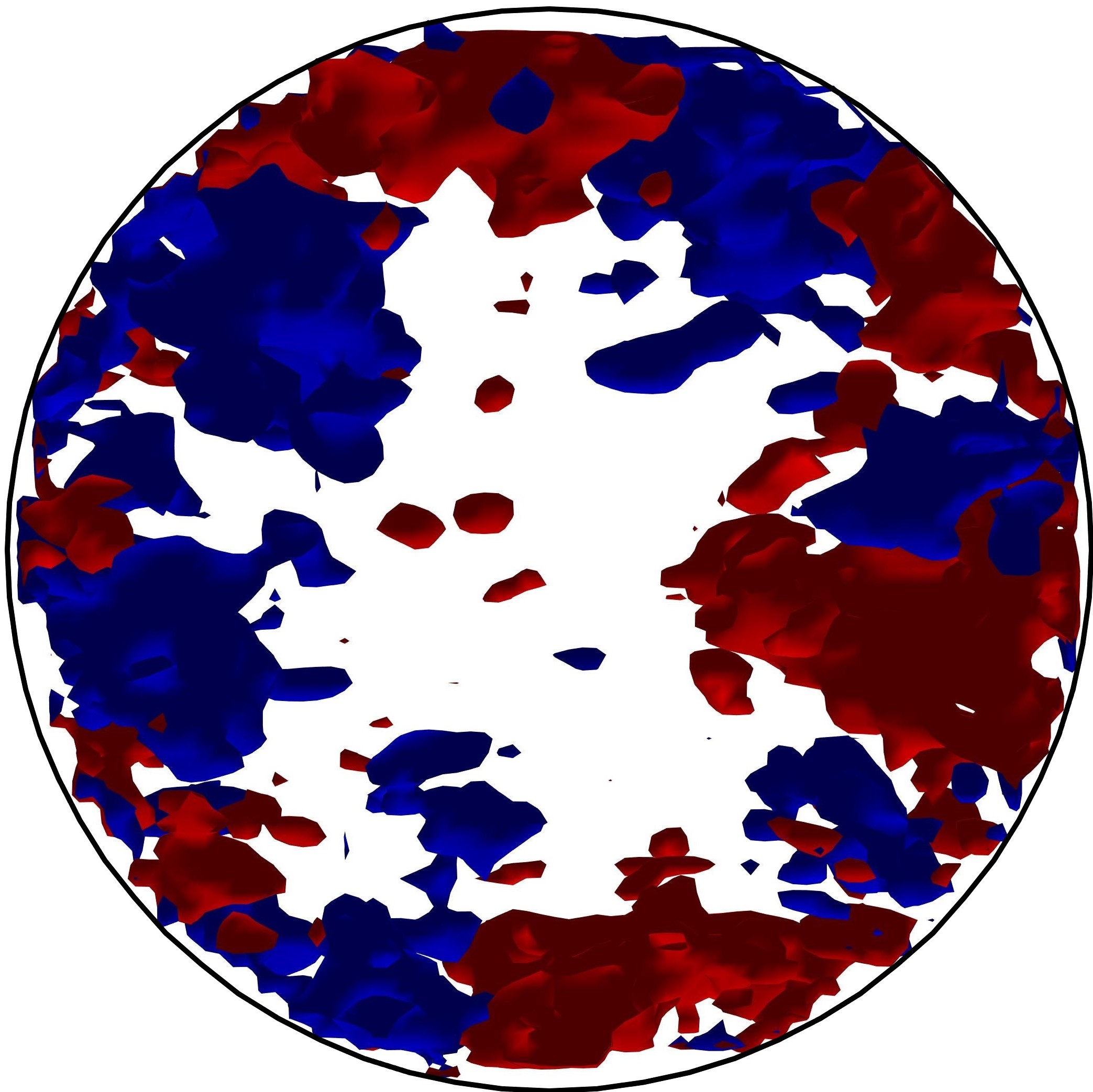}}
        \caption{$~k_{\theta}=6$}
    \end{subfigure}
        \hfill
\begin{subfigure}[t]{0.26\textwidth}
        \raisebox{-\height}{\includegraphics[width=\textwidth]{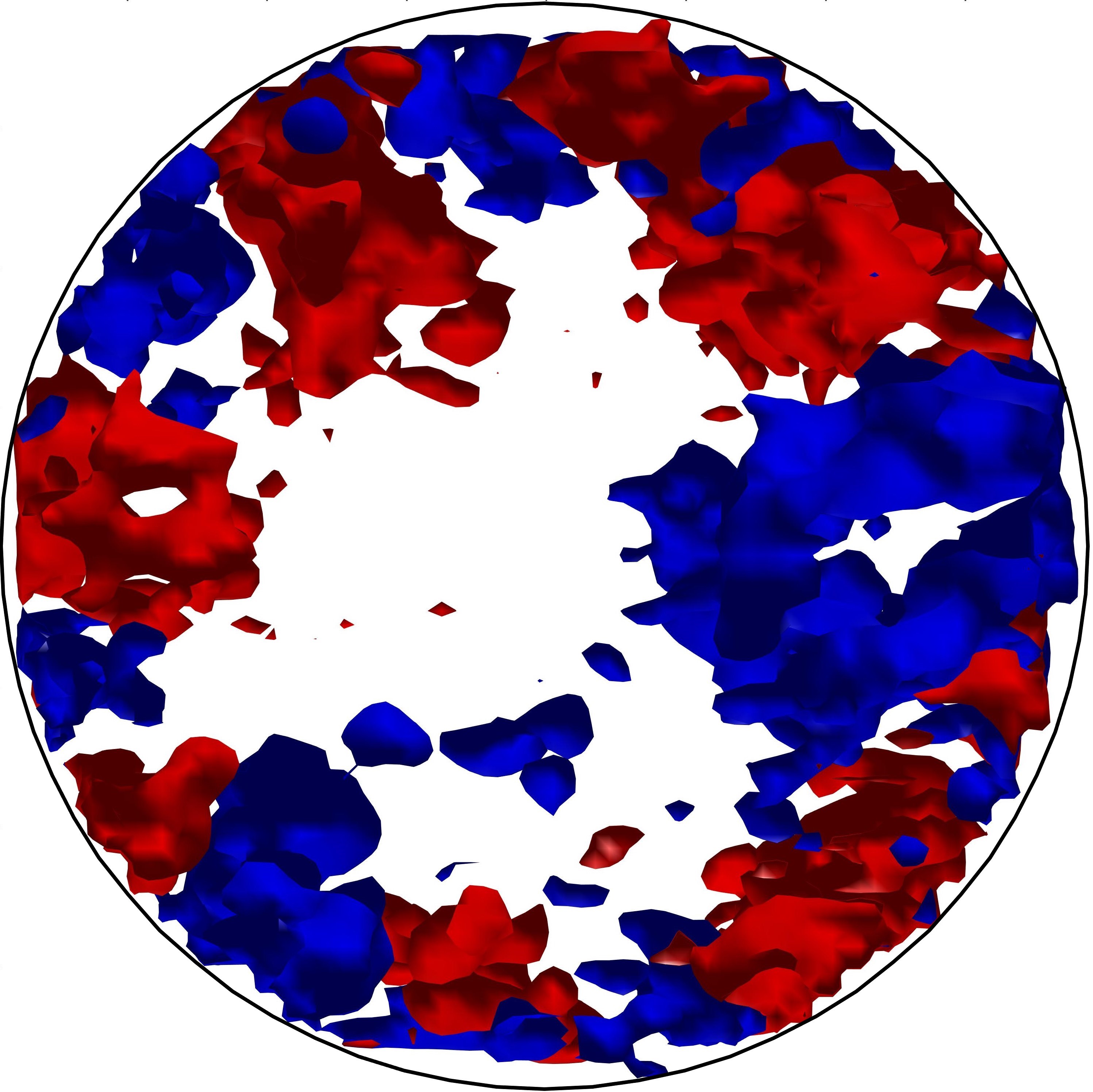}}
        \caption{$~k_{\theta}=7$}
    \end{subfigure}

\caption{\label{fig:states_inst_24400} Instantaneous snapshots of velocity field fluctuations assigned to azimuthal wave numbers 2 to 7 of Re =24400. The red and blue iso-contours correspond to velocities 1.5 per cent above and below the mean velocity profile, respectively.}
 \label{fig:three graphs}
\end{figure*}

Fig. \ref{fig:states_inst_24400} presents examples for instantaneous snapshots of velocity field fluctuations of Re =24400 assigned to azimuthal the wave numbers 2 to 7 by our detection method. Although small-scale fluctuations govern the background, the dominant pattern of streaks along the azimuth clearly resembles its respective wave number.



\begin{figure*}
     \centering
    \begin{subfigure}[t]{0.26\textwidth}
        \raisebox{-\height}{\includegraphics[width=\textwidth]{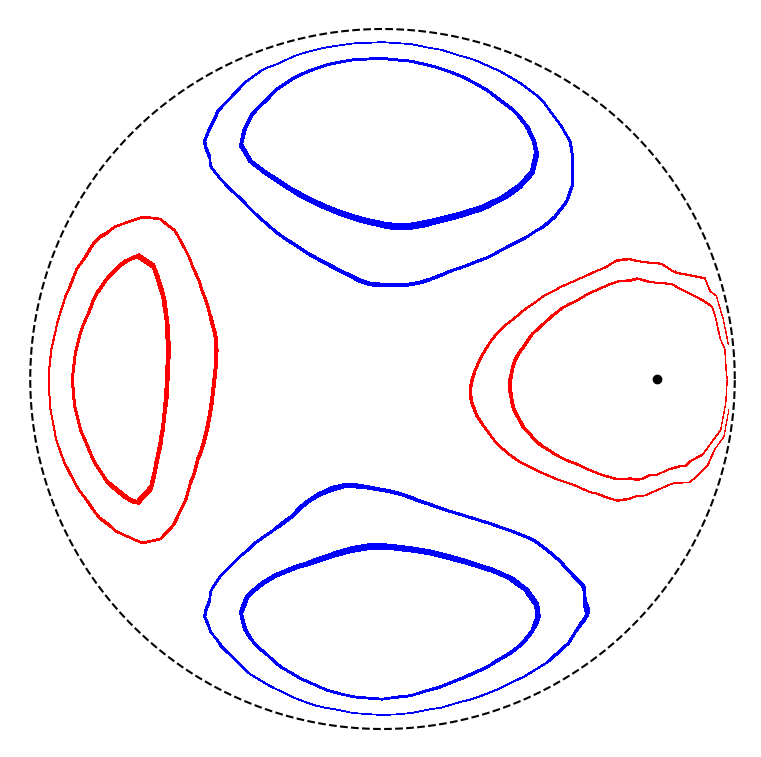}}
        \caption{$~k_{\theta}=2$}
    \end{subfigure}
    \hfill
    \begin{subfigure}[t]{0.26\textwidth}
        \raisebox{-\height}{\includegraphics[width=\textwidth]{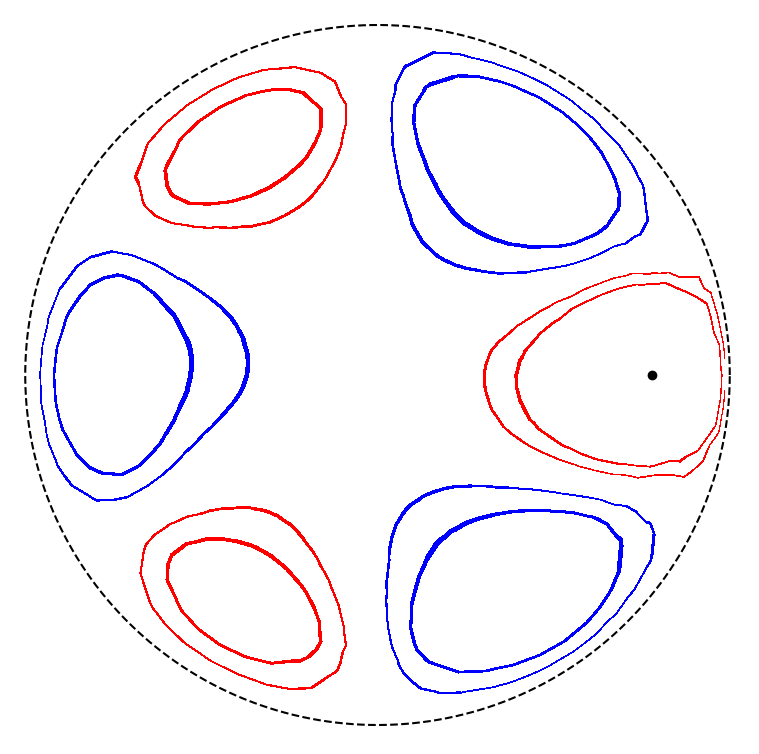}}
        \caption{$~k_{\theta}=3$}
    \end{subfigure}
        \hfill
\begin{subfigure}[t]{0.26\textwidth}
        \raisebox{-\height}{\includegraphics[width=\textwidth]{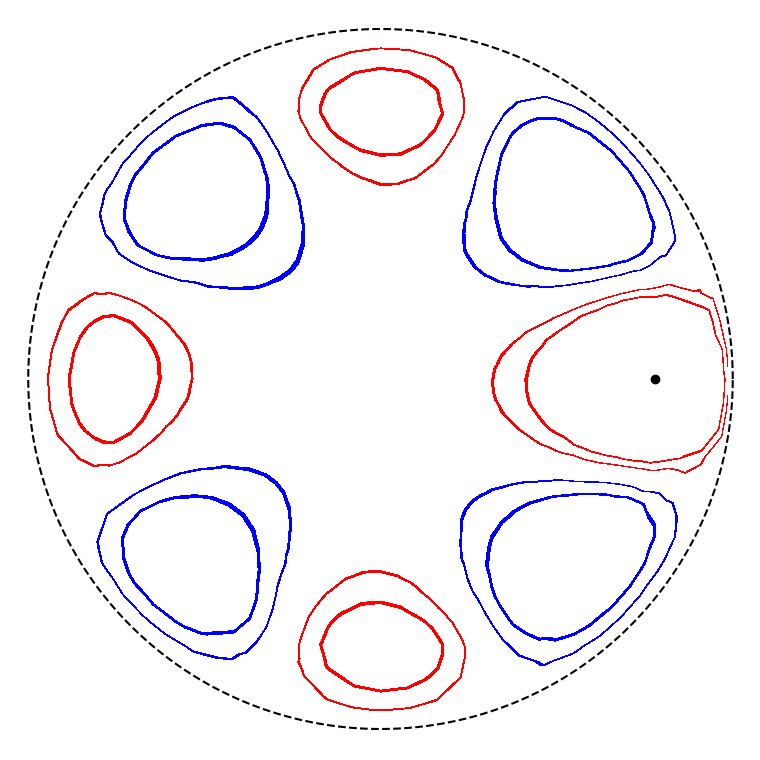}}
        \caption{$~k_{\theta}=4$}
    \end{subfigure}

\vspace{0.5cm}
    
\begin{subfigure}[t]{0.26\textwidth}
        \raisebox{-\height}{\includegraphics[width=\textwidth]{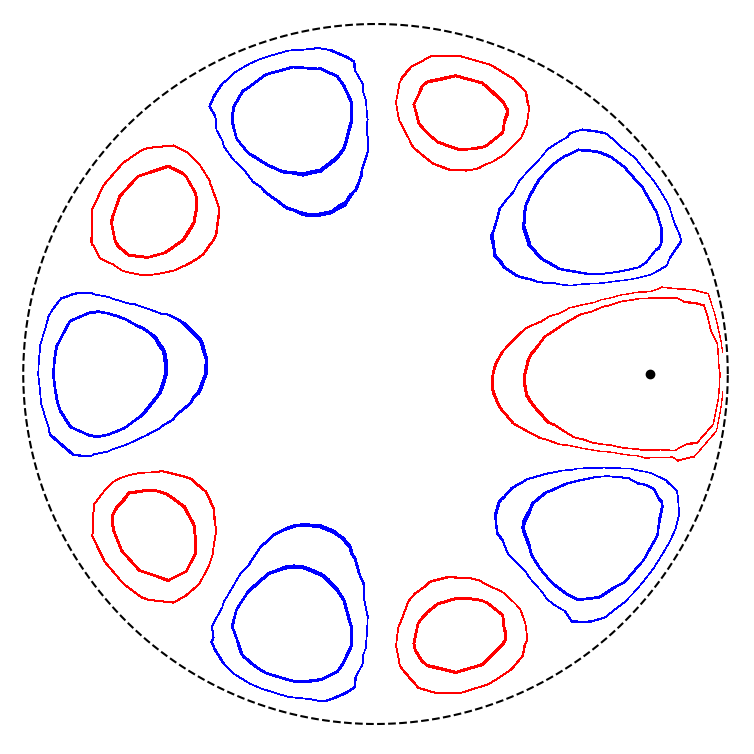}}
        \caption{$~k_{\theta}=5$}
    \end{subfigure}
    \hfill
    \begin{subfigure}[t]{0.26\textwidth}
        \raisebox{-\height}{\includegraphics[width=\textwidth]{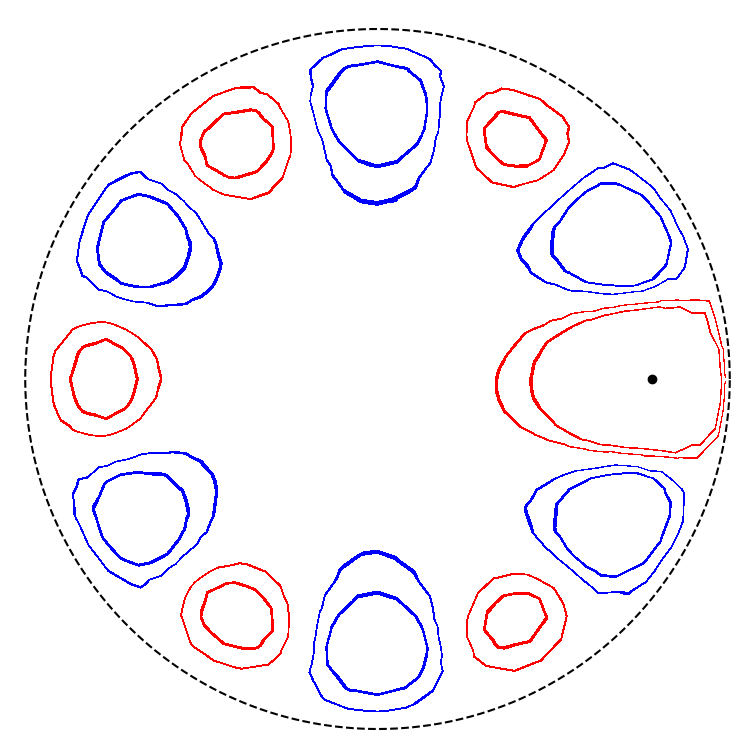}}
        \caption{$~k_{\theta}=6$}
    \end{subfigure}
        \hfill
\begin{subfigure}[t]{0.26\textwidth}
        \raisebox{-\height}{\includegraphics[width=\textwidth]{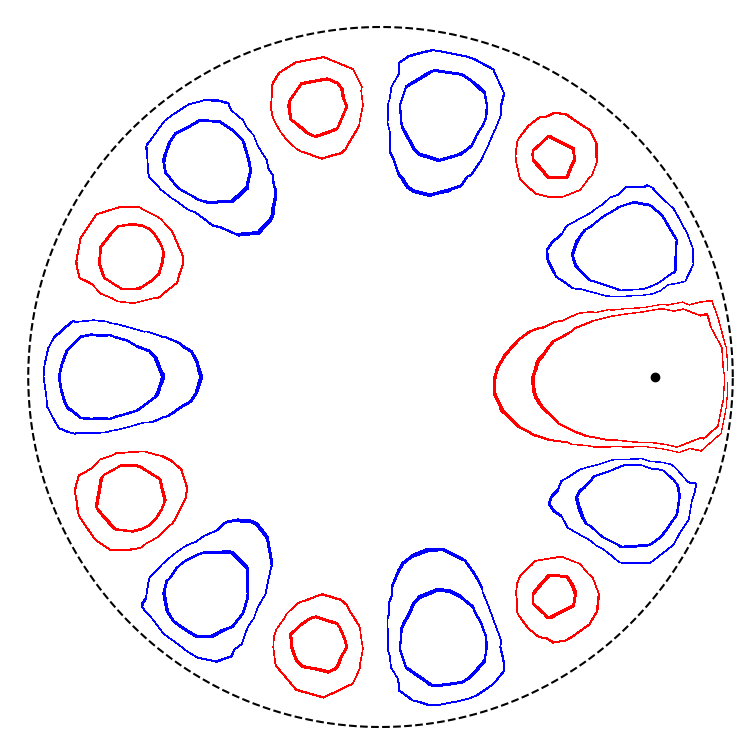}}
        \caption{$~k_{\theta}=7$}
    \end{subfigure}

\caption{\label{fig:states_4900} Organizational turbulent states with azimuthal wave numbers of 2 to 7 of Re =17800. The images show the spatial correlation function $R_{uu}$, the red level curves correspond to $R_{uu}$ = 0.05 and 0.1, and the blue ones to the opposite sign. The reference point is illustrated as a black dot. }
 \label{fig:three graphs}
\end{figure*}

Fig. \ref{fig:states_4900} exemplarily visualizes the correlation contour levels of the coherent wave number states 2 to 7 obtained by the conditional averaging procedure for the Reynolds number of Re = 17800. The spatial correlation $R_{uu}$ is visualized by iso-contours with respect to the reference point, whose location was also visualized by a black dot. The red level curves correspond to $R_{uu}$ = 0.05 and 0.1, and the blue ones to the opposite sign.

\begin{figure*}
\centering
\includegraphics[width=1\textwidth]{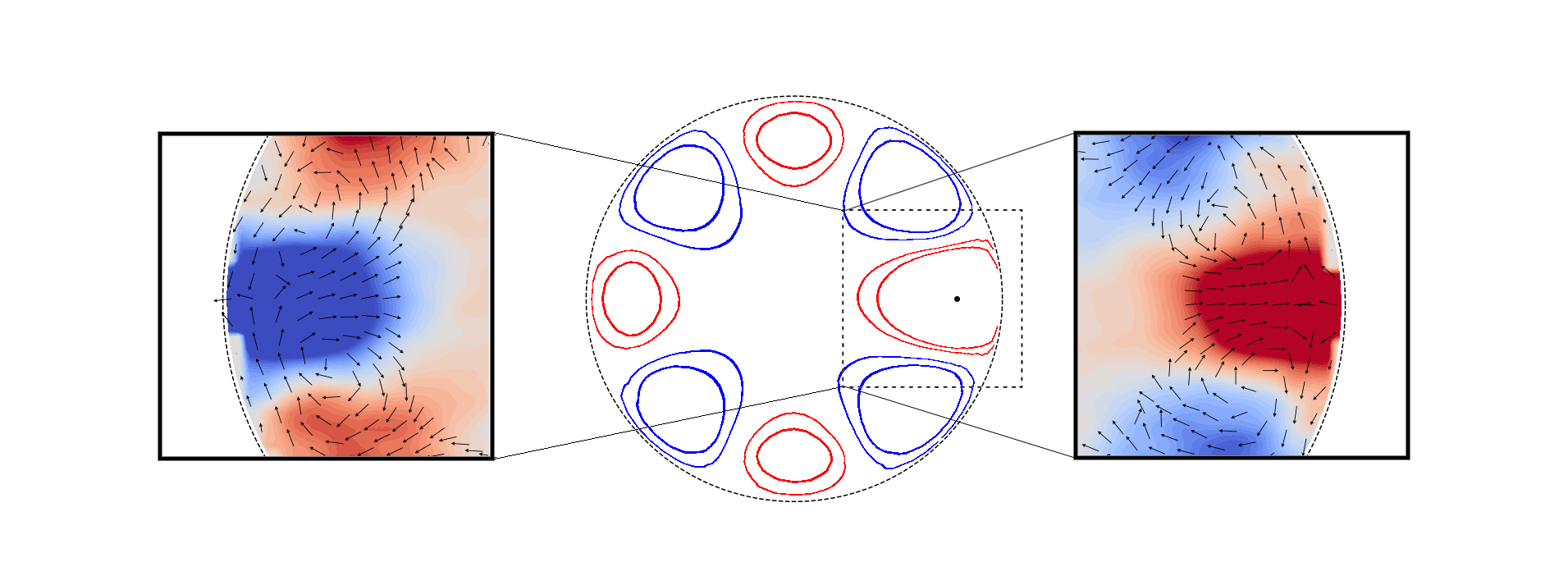}
\caption{\label{fig:vector_field} Conditionally averaged vector field (left) and corresponding principal coherent patterns (right) related to regions of positive (red) and negative velocity fluctuations (blue) of wave number state 4 of Re = 17800.}

\end{figure*}

By means of conditional averaging, we visualized the coherent flow patterns in the cross-stream slice in Fig. \ref{fig:vector_field} for the wave number state 4 of Re = 17800. All in-plane vectors were normalized to the same magnitude in order to improve the visualization, particularly of the vortex patterns.



\noindent
For all states, we were able to identify this alternating pattern of streamwise fluctuations, similar to the one observed by Dennis and Sogaro \cite{Dennis2008}. On average, the areas of negative streamwise fluctuations appear to extend more towards the pipe center than the positive. For all conditionally averaged vector fields, we clearly observe that regions of positive streamwise fluctuation are related to an in-plane movement towards the pipe wall, while negative regions of streamwise fluctuations show a movement in the opposite direction, towards the pipe centre. These strong radial motions are accompanied by pairs of weaker, counter-rotating vortices that are saddled symmetrically along the lateral sides of the fluctuation regions. They are related to the shear layer between the regions of opposed radial motions.
The ability to observe well-defined in-plane patterns underlines the potential of conditional averaging to decipher the apparently chaotic nature of turbulent flow fields, having in mind that an unconditional average of all flow field samples would just zero out all vector components apart from the mean flow direction. The principal coherent patterns that we found to govern the cross-stream vector fields of both Reynolds numbers are illustrated in Fig. \ref{fig:vector_field}. The relation between the streamwise and radial velocity components coincides with earlier findings in the quadrant analysis in turbulent pipe flow \cite{Brodkey1969}, showing the dominance of $Q_{2}$ and $Q_{4}$ quadrant motions in the near-wall region.

\begin{figure}
\includegraphics[width=0.49\textwidth]{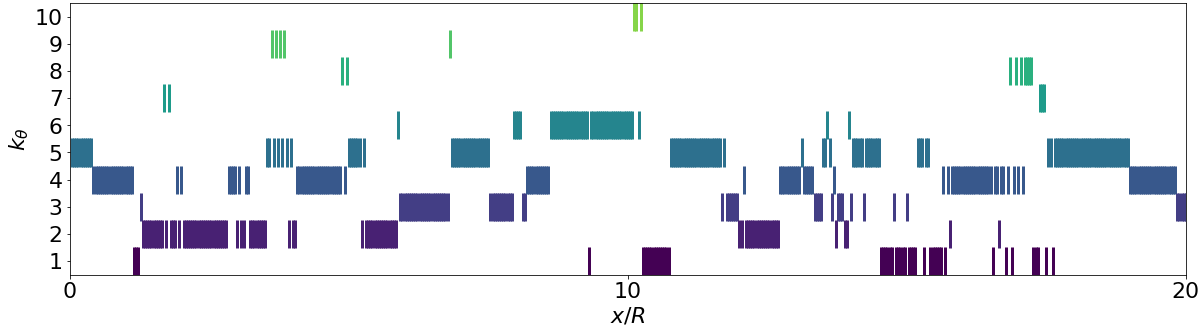}
\caption{\label{fig:streamswise_extent_bars_5300} Streamwise extent of coherent states along the main flow direction in an arbitrary section 20 pipe radii at Re = 5300.}
\end{figure}

By the application of Taylor's hypothesis, Fig. \ref{fig:streamswise_extent_bars_5300} exemplarily illustrates the advection of wave number states and their corresponding spatial extent along the main flow direction along an arbitrary section of 20 pipe radii for Re = 5300. At first glance, several wave number structures with streamwise extent within the order of the pipe radius can be identified. Longer structures can be observed for the wavenumbers 2 to 6. The remaining wave number states, on the other hand, show structures of a more intermittent nature, including several one-snapshot observations.

\begin{figure}
\includegraphics[width=0.49\textwidth]{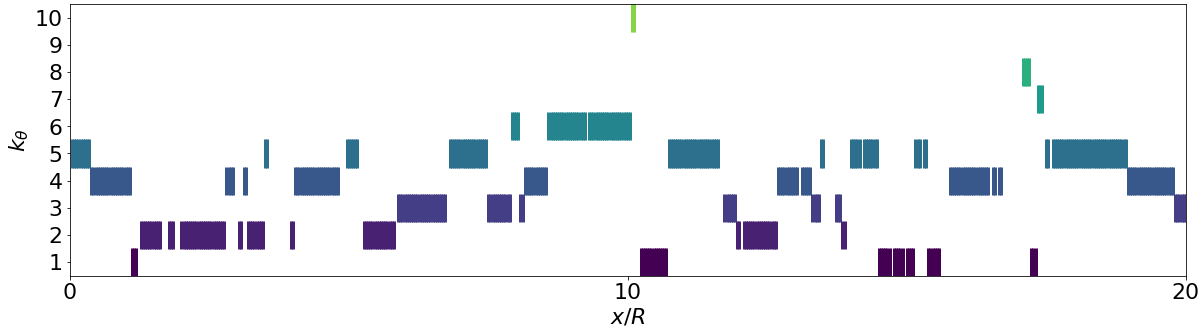}
\caption{\label{fig:streamswise_extent_bars_5300_without_ones} Streamwise extent of coherent states along the main flow direction in an arbitrary section 20 pipe radii at Re = 5300 without unstable remainders.}
\end{figure}

Because we were particularly interested in structures with a streamwise extent, in Fig. \ref{fig:streamswise_extent_bars_5300_without_ones} we excluded the wave number snapshots that were only observed in one snapshot. 
On these unstable wave number observations -hereafter called unstable remainders- we will take a closer look in the next section.

\subsubsection{\label{sec:level3}Statistical distribution of coherent states }

\begin{figure}
\includegraphics[width=0.45\textwidth]{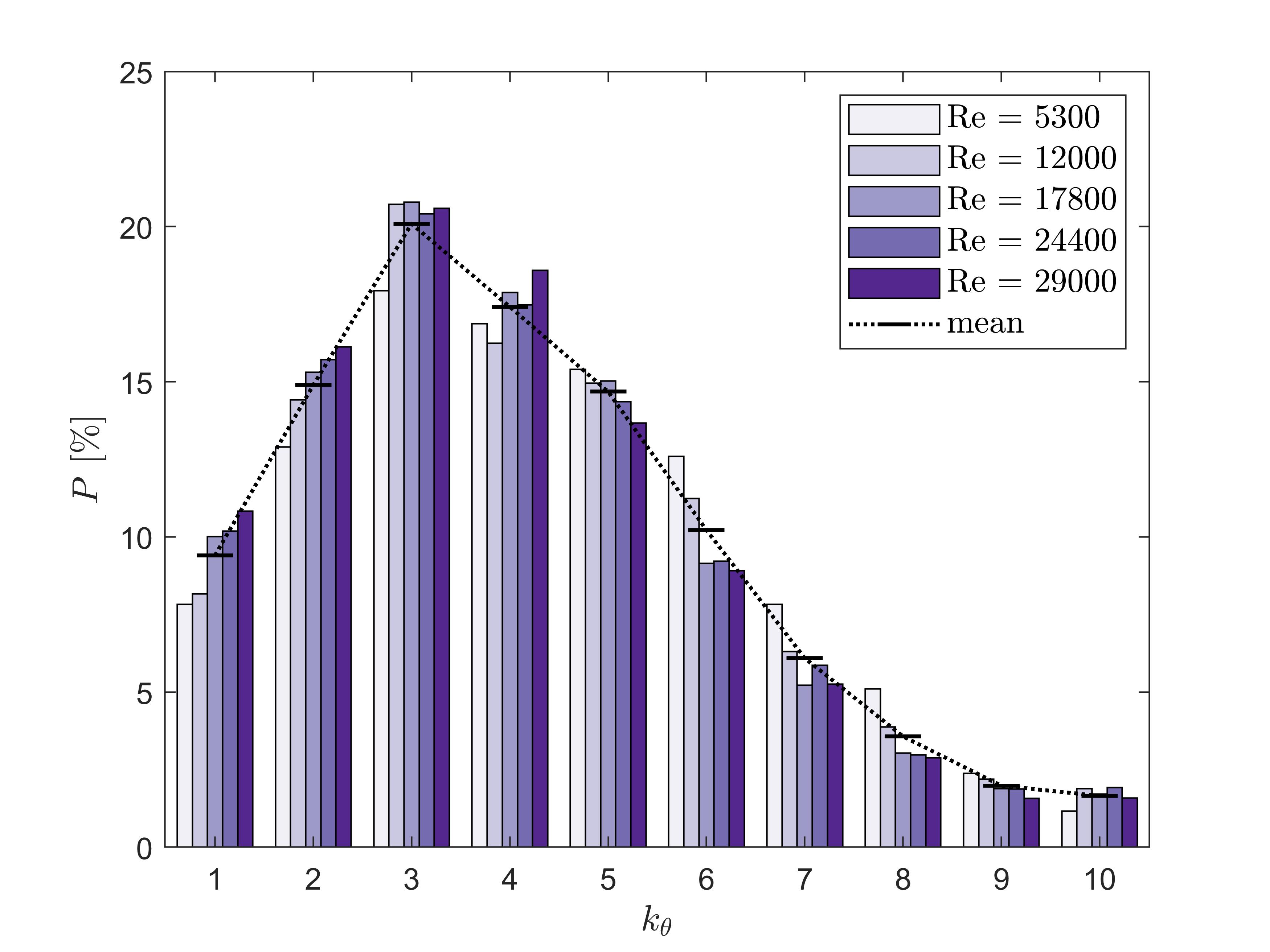}
\caption{\label{fig:weight_distribution} Normalized distribution of dominant wave number states based on snapshot observations.}
\end{figure}

As emphasized in the foregoing, we applied a second conditional average on the allocated wave number vector fields with respect to the sign of the streamwise fluctuation in the reference point. The corresponding snapshot proportions are well-balanced for the states of all Reynolds numbers. We take this observation as an indicator that the number of samples allocated to each state is sufficient to consider our results as statistically converged for the ten wave number states we present. 
We first present the statistical weight distributions, i.e. the percentual contribution of the ten dominant states with respect to the total number of state-assigned vector fields in Fig. \ref{fig:weight_distribution}, without applying any threshold.

We observe that for all  Reynolds numbers the weight distributions show positive skewness towards lower states. For all Reynolds numbers, the most detected state was wave number 3, which is in agreement with the previous observations \cite{Dennis2008} for Re = 35000 and the most energetic azimuthal mode found using Proper Orthogonal Decomposition (POD) of turbulent pipe flow at Re = 24580 obtained with Direct Numerical Simulation (DNS) by Baltzer el al. \cite{Baltzer2013}.

Of the ten dominant wave numbers, state 10 showed the lowest statistical weight. We also detected wave number states above 10, but their statistical contribution was very low (below 1 per cent in all the sets) and might not show converged statistics. The highest wave numbers we detected in each set showed to increase with growing Reynolds number, namely 13, 14, 16, 21 and 22 for the Reynolds numbers of 5300, 12000, 17800, 24400, and 29000, respectively.

\begin{figure}
\includegraphics[width=0.45\textwidth]{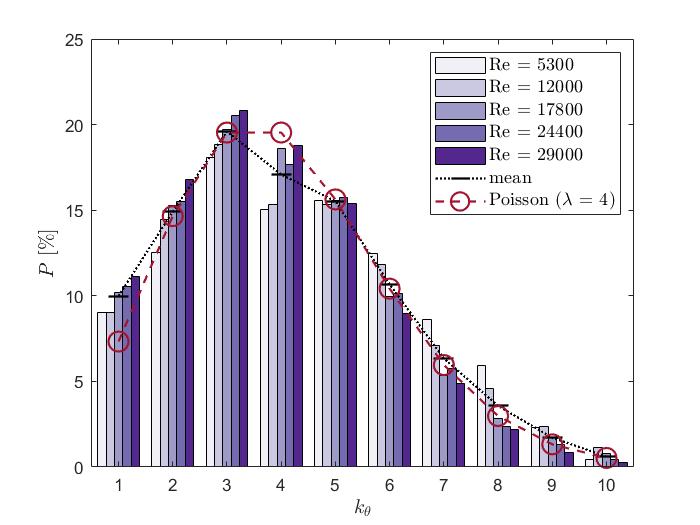}
\caption{\label{fig:weight_distribution_structures}Normalized distribution of dominant wave number states based on structures with a streamwise extent.}
\end{figure}

Figure \ref{fig:weight_distribution_structures} shows the normalized distribution from the perspective of wave number structures with a streamwise extent. The unique feature of this presentation is that the statistical weight distribution is calculated with respect to the number of structures (independent of the number of snapshots it consists of) passing through the measurement plane, and not solely on the raw number of snapshots assigned to the wave number sets. 

Because we were interested in structures with a streamwise extent, we excluded the vector fields that were only observed in one single snapshot, namely the unstable remainders, from this representation. Nevertheless, the unstable remainders vector fields showed a significant weight contribution, namely 25.2, 20.9, 28.1, 33.3, and 35.3 per cent for the Reynolds number sets of 5300, 12000,17800, 24400, and 29000, respectively. 
With respected to the spatial resolution of the unstable remainders note that the advection velocity of the structures is increasing with the Reynolds number although the sampling rate was held constant (constrained by the maximum laser frequency) for all Reynolds number sets. This implies a different advected structure length between two snapshots for each Reynolds number, namely 0.028 pipe radii for the lowest Re of 5300 and 0.148 pipe radii for the highest Re of 29000.

Comparing Figs. \ref{fig:weight_distribution} and \ref{fig:weight_distribution_structures}, we see that from the structure's perspective, the weight contribution of higher wave number structures is generally lower than observed from the viewpoint of snapshots. We interpret this as an indicator that higher wave number states present a more intermittent behavior, often being cut off as unstable remainder states in the structure representation.

Another interesting statistical feature is that the distribution of wave number structures is reasonably well-described by a Poisson distribution with a mean value of $\lambda$ = 4 (see Fig. \ref{fig:weight_distribution_structures}). Based on the nature of Poisson distributions, this is a hint that the transition to a new wave number structure is independent of the present state and supports earlier assumptions of an underlying Markovian description \cite{Jäckel_2023}.

\noindent

For both the snapshot's and the structure's perspective, we observe that the weight distribution of states is generally independent of the Reynolds number. 
On a closer perspective, we observe a very slight shift to the right with increasing Reynolds number, i.e., higher wave number states become more frequent. 
Schneider et al. \cite{Schneider2007} also observed that their weight distribution of states shifts to the right with increasing Reynolds number. Nevertheless, their work is focused on transitional pipe flow and also their window of observation was relatively small (Re = 2200, 2350, and 2500) for a strong statement on behalf of that matter. Furthermore, their state detection incorporates a cut-off threshold which is very likely the reason for the overall low weight contributions in the occurrence statistics of their wave numbers. Even with their relatively big data sets (ca. 15000- 17000 vector fields) it is difficult to state if their state contributions are statistically converged, particularly for the less encountered states. The generally constant weight distribution of wave number states for all Reynolds number leads us to the statement that the aforementioned Reynolds number effect, i.e. the Reynolds number dependence of turbulent statistics in the range of moderate Reynolds numbers, is not reflected from the dynamical systems viewpoint of coherent states.

\section{Discussion}

We set up a 6-inch diameter pipe flow loop with a SPIV system to investigate a number of interesting open issues related to coherent states in turbulent pipe flows. A robust detection algorithm was developed which is not affected by the background fluctuation of the flow. With this setup, we were able to reveal the nature of these states by visualizing their inherent patterns and uncovering interesting statistical features. In this way, we closed the huge gap between the Reynolds numbers at which Schneider et al. \cite{Schneider2007} and Dennis and Sogaro \cite{Dennis2008} observed these structures.   
Our key observations are presented in the following: 	 
\begin{enumerate}
   
  \item For all investigated Reynolds numbers, 10 dominant states were identified, consisting of patterns of alternating streaks along the pipe’s azimuth. We thereby confirm Dennis and Sogaro’s \cite{Dennis2008} assertion, that coherent states are not phenomena of laminar-turbulent transition as assumed earlier, but govern also the dynamics of fully developed turbulent pipe flow.  
  
  \item  For all investigated Reynolds numbers, the weight distribution of states shows probabilities with positive skewness towards lower states, with a most encountered azimuthal wave number of 3. The weight distribution of wave number structures is reasonably-well described by a Poisson distribution. This is a hint that the transition to a new wave number structure is independent of the present state.
  
  \item The weight distribution of wave number states for all Reynolds number is very similar. We conclude that the Reynolds number effect, i.e. the Reynolds number dependence of turbulent statistics in the range of moderate Reynolds numbers, is not reflected from the dynamical systems viewpoint of coherent states.

\end{enumerate}

There is a lot of future work required in this new field of turbulence: From a phenomenological perspective, we are particularly interested if it is possible to uncover any Reynolds number dependencies. Therefore, more Reynolds number sets are currently being measured in order to increase the resolution within the range of moderate Reynolds number flow and thereby increase our sensibility to unveil possible tendencies.
In parallel, we are currently processing the present data sets to present more statistical features regarding the streamwise organization of the wave number states, \textit{inter alia} the maximum and average length distributions, as well as their state recurrence timescales. 
We assume that coherent states, apart from their phenomenological importance for the understanding of the nature of turbulence \cite{Moriconi_2009}, play a key role in some of the most relevant fields of fluid engineering, e.g. as contributors to the Reynolds stresses, as well as to the heat- and species transport between the bulk and near-wall region. 
Coherent motions likely have a strong contribution to high particle concentrations close to the wall, namely turbophoresis, which causes to scaling of pipe walls, one of the key issues to be tackled in particle-laden flows.  Therefore, we are interested in mechanisms to control the coherent structures. For instance, we address how the magnetic fields can influence coherent motions by turbulent dissipation from a theoretical, experimental and numerical perspective \cite{Moriconi2020, Moriconi2022,Tavares_2022, Magacho_2023}, and suspect that damping of the coherent sweeps can reduce an important chain element of the process that transports scaling particles close to the wall region. 


\begin{acknowledgments}

This research received financial support from the Brazilian National Council for Scientific and Technological Development (CNPq) and Petrobras. The authors gratefully acknowledge this support.
\end{acknowledgments}

\section*{Data Availability Statement}

The data that support the findings of this study are available from the corresponding author upon reasonable request.

\appendix

\section{\label{app:Appendix}Detection- and visualization procedure}

\begin{figure}
\includegraphics[width=0.5\textwidth]{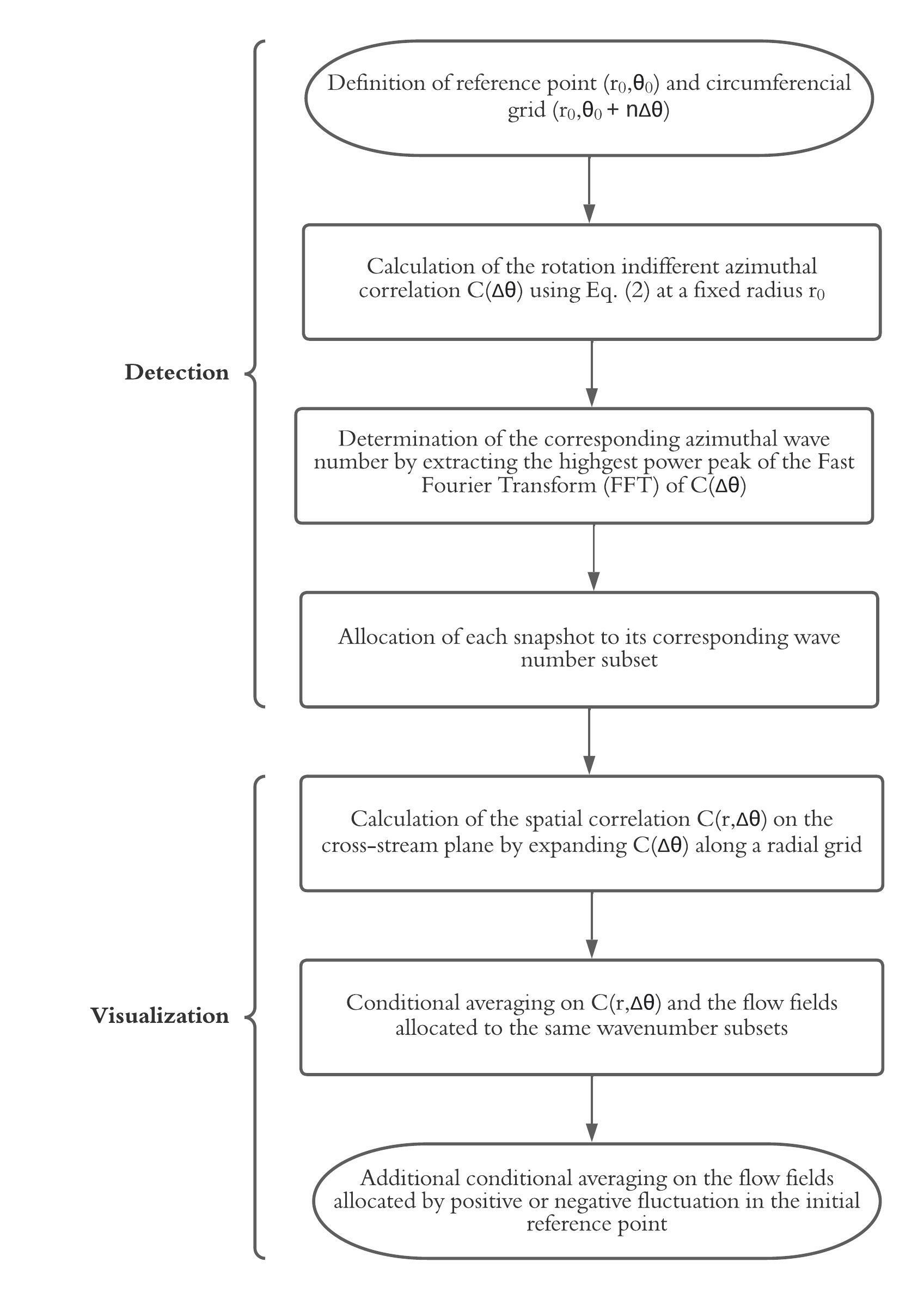}
\caption{\label{fig:principal_steps} Principal steps for the detection and visualization of coherent states.}
\end{figure}

In Fig. \ref{fig:principal_steps}, we present a concise illustration of the methodology for the detection and visualization of the coherent states. First, in the detection procedure,  by means of a spatial correlation function long the pipe's azimuth, the individual snapshot's wave number can be obtained by an FFT- analysis. Its inherent flow field is then allocated to its corresponding state bin. Then, in the visualization procedure, the coherent patterns are reconstructed by means of conditional averaging of the state-assigned flow fields. 

\nocite{*}
\bibliography{aipsamp}

\end{document}